\documentclass[aps,pre,superscriptaddress,showkeys,citeautoscript]{revtex4-2}
\usepackage[colorlinks=true,bookmarks=false,citecolor=blue,linkcolor=red,hyperfootnotes=true,urlcolor=blue]{hyperref}

\usepackage{physics}
\usepackage{graphicx}% Include figure files
\usepackage{dcolumn}% Align table columns on decimal point
\usepackage{bm}% bold math
\usepackage[dvipsnames]{xcolor}
\usepackage{wasysym}
\usepackage{lipsum}
\usepackage{enumitem}
\usepackage{booktabs} 
\usepackage{gensymb}
\usepackage{xcolor}
\usepackage{amssymb}
\usepackage{booktabs}
\usepackage{array}
\newcolumntype{L}{>{$}l<{$}}
\newcolumntype{C}{>{$}c<{$}}
\newcolumntype{R}{>{$}r<{$}}

\usepackage{blindtext}

\makeatletter
\def\p@subsection{}
\makeatother

\newcommand{\code}[1]{\texttt{#1}}

\begin{document}

\title{Lightshow: a Python package for generating computational x-ray absorption spectroscopy input files}

\author{Matthew R. Carbone}\email{mcarbone@bnl.gov}
\thanks{Equally-contributing author}
\affiliation{Computational Science Initiative, Brookhaven National Laboratory, Upton, New York 11973, United States}

\author{Fanchen Meng}\email{fmeng1@bnl.gov}
\thanks{Equally-contributing author}
\affiliation{Center for Functional Nanomaterials, Brookhaven National Laboratory, Upton, New York 11973, United States}

\author{Christian Vorwerk}
\affiliation{Pritzker School of Molecular Engineering, University of Chicago, Chicago, Illinois 60637, United States}

\author{Benedikt Maurer}
\affiliation{Physics Department and IRIS Adlershof, Humboldt-Universität zu Berlin, D-12489 Berlin, Germany}

\author{Fabian Peschel}
\affiliation{Physics Department and IRIS Adlershof, Humboldt-Universität zu Berlin, D-12489 Berlin, Germany}

\author{Xiaohui Qu}
\affiliation{Center for Functional Nanomaterials, Brookhaven National Laboratory, Upton, New York 11973, United States}

\author{Eli Stavitski}
\affiliation{National Synchrotron Light Source II, Brookhaven National Laboratory, Upton, New York 11973, United States}

\author{Claudia Draxl}
\affiliation{Physics Department and IRIS Adlershof, Humboldt-Universität zu Berlin, D-12489 Berlin, Germany}

\author{John Vinson}
\affiliation{Material Measurement Laboratory, National Institute of Standards and Technology, Gaithersburg, Maryland 20899, United States}

\author{Deyu Lu}
\affiliation{Center for Functional Nanomaterials, Brookhaven National Laboratory, Upton, New York 11973, United States}

\date{\today}

\begin{abstract}
First-principles computational spectroscopy is a critical tool for interpreting experiment, performing structure refinement, and developing new physical understanding. Systematically setting up input files for different simulation codes and a diverse class of materials is a challenging task with a very high barrier-to-entry, given the complexities and nuances of each individual simulation package. This task is non-trivial even for experts in the electronic structure field and nearly formidable for non-expert researchers. \code{Lightshow} solves this problem by providing a uniform abstraction for writing computational x-ray spectroscopy input files for multiple popular codes, including FEFF, VASP, {\sc ocean}, \texttt{exciting} and {\sc xspectra}. Its extendable framework will also allow the community to easily add new functions and 
to incorporate new simulation codes.
\end{abstract}


\maketitle

\section{Statement of need}

First-principles simulations explore materials and molecular properties of a system by solving fundamental quantum mechanical equations numerically. Thanks to their predictive nature, first-principles simulations provide fundamental understanding into the physical origins of various phenomena at the microscopic level, making them a powerful tool at the forefront of a wide range of scientific research fields, including physics, chemistry, materials science and biology. They are also critical to accelerating new materials design. In comparison to experiments, which can be expensive and time-consuming, \textit{in silico} materials design frameworks can quickly screen the most promising candidates for target applications by running high-throughput calculations, allowing for a systematic down-sampling of the intractably large chemical space. The emergence of the high-performance computing hardware architecture combined with the development of efficient structure search algorithms continue to fuel the advance of first-principles simulations in materials design.

Spectroscopy is an important experimental characterization technique that probes a sample based on the physics of light-matter interaction. Different types of spectroscopy can be classified by the energy range they probe, such as X-ray, ultraviolet–visible, and infrared spectroscopy. \code{Lightshow} currently focuses on writing the input files for one type: X-ray absorption spectroscopy (XAS), in which a deeply bound core level electron is excited to empty states in the  conduction bands. XAS is particularly useful because it is element-specific and very sensitive to the local chemical environment of the absorbing sites, such as coordination number, charge state, and local symmetry~\cite{de2008core}. It has been widely used in condensed matter physics, geophysics, chemistry, materials science and biology for materials characterization. Recent instrument development at synchrotron light sources further improves the spatial, temporal and energy resolution of XAS, which opens new avenues in XAS research.

Despite the growing demands for first-principles XAS spectroscopy, carrying out practical calculations correctly is far from trivial and requires a great deal of expertise in electronic structure theory, creating a formidable barrier for non-expert researchers. Most of the practical challenges boil down to the proper choice of input parameters, which depends on the level of theory, details of the implementation of the simulation software, and the atomic structure of the system. A general purpose software package for generating XAS simulation input files for multiple codes does not exist. \code{Lightshow} has been developed to fill this gap. It provides not only sets of default input parameters based on a careful multi-code XAS benchmark project~\cite{xanesbench}, but also exposes the entire suite of possible parameter choices to expert users to tune. Our goal is to provide an easy-to-use tool to the XAS community (for both newcomers and experts) for XAS simulation and analysis. This tool will also help to improve the data consistency and data reproducibility in the computational x-ray spectroscopy field,  which are essential to data-driven applications.

\begin{figure}
    \centering
    \includegraphics[width=\columnwidth]{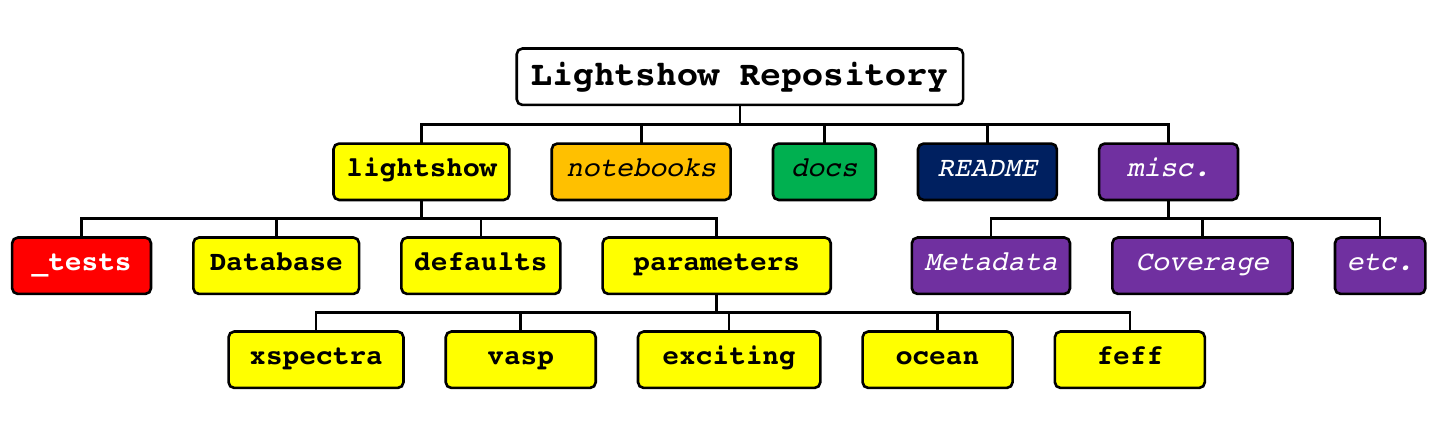}
    \caption{
        Graphical representation of the organization of the \code{Lightshow} repository. \label{fig:WorkflowDiagram}
    }
\end{figure}

\section{Brief software description}

We summarize the structure of \code{Lightshow}'s application programming interface (API) in Fig.~\ref{fig:WorkflowDiagram}. \code{Lightshow}'s core design philosophy is built around two principal objects: the \code{Database} class and the \code{\_BaseParameters} class. At a high level, the \code{Database} class  interfaces primarily with Pymatgen and the Materials Project~\cite{Jain2013},  allowing the user to easily utilize Pymatgen and pull a large number of materials  structures quickly. Functionality is also available for instantiating a  \code{Database} via loading, e.g., \code{POSCAR}-style structure files from local  storage. Once a database has been created, code-specific simulation parameters  inheriting the \code{\_BaseParameters} base class interface with various methods in Pymatgen as well as in-house built software for systematically writing input files for multiple XAS simulation programs, including FEFF~\cite{rehr2010parameter}, {\sc xspectra}~\cite{taillefumier2002x,gougoussis2009first,bunuau2013projector}, {\sc ocean}~\cite{vinson2011bethe,ocean-3}, \texttt{exciting}~\cite{exciting} and VASP~\cite{vasp-xas}. We highlight the \code{\_tests} directory, in which we maintain a suite of unit tests for individual functions and methods, as well as integration tests for the entire workflow. \code{Lightshow} is fully documented, and contains a simple example notebook for users to get started.

\section{Concluding notes}
The Lightshow package can be found open source under a BSD-3-clause license on GitHub~\footnote{\href{https://github.com/AI-multimodal/Lightshow}{github.com/AI-multimodal/Lightshow}}, or can be installed via pip using \code{pip install lightshow}.

\begin{acknowledgements}
This work is partially supported by the U.S. Department of
Energy, Office of Science, Office of Basic Energy Sciences, under Award Numbers
FWP PS-030. The research used the theory and computation resources of the
Center for Functional Nanomaterials, which is a U.S. DOE Office of Science
Facility, and the Scientific Data and Computing Center, a component of the
Computational Science Initiative, at Brookhaven National Laboratory under
Contract No. DE-SC0012704. This work received partial funding by the German
Research Foundation (DFG) through the CRC 1404 (FONDA), Projektnummer
414984028, and the NFDI consortium FAIRmat – project 460197019.

C.V. acknowledges support by the Department of Energy, Basic Energy Sciences,
Materials Science and Engineering Division, through the Midwest Integrated
Center for Computational Materials (MICCoM).

Certain software is identified in this paper for clarity. Such identification
is not intended to imply recommendation or endorsement by NIST, nor is it
intended to imply that the materials or equipment identified are necessarily
the best available for the purpose.
\end{acknowledgements}

\providecommand{\noopsort}[1]{}\providecommand{\singleletter}[1]{#1}%

\end{document}